\documentclass[dvipdfmx,letterpaper,conference]{ieeeconf}
\IEEEoverridecommandlockouts
\usepackage{graphicx}
\usepackage{amsmath,amssymb}
\newtheorem{theorem}{Theorem}
\newtheorem{definition}{Definition}

\newtheorem{lemma}{Lemma}
\newtheorem{remark}{Remark}
\newtheorem{corollary}{Corollary}
\newtheorem{proposition}{Proposition}

\makeatletter
\def\IEEEproofname{Proof}
\def\IEEEproof{\@ifnextchar[{\@IEEEproof}{\@IEEEproof[\IEEEproofname]}}
\def\@IEEEproof[#1]{\par\noindent\hspace{2em}{\itshape #1: }}

\makeatother

\DeclareMathOperator{\diag}{diag}
\DeclareMathOperator{\tr}{tr}

\usepackage{physics}
\usepackage{cite}
\usepackage{mathtools}
\mathtoolsset{showonlyrefs=true}
\usepackage{mathrsfs}
\usepackage{bm}
\usepackage{optidef}

\usepackage{algorithm}
\usepackage{algorithmicx}
\usepackage{algpseudocode}
\usepackage{flushend}
\usepackage{tikz}
\usetikzlibrary{calc}
\usetikzlibrary{positioning}
\usetikzlibrary{arrows.meta}

\newcommand{\bbR}{\mathbb{R}}

\newcommand{\bbZ}{\mathbb{Z}_+}

\newcommand{\scrS}{\mathscr{S}}
\newcommand{\scrT}{\mathscr{T}}

\newcommand{\scrA}{\mathscr{A}}
\newcommand{\scrB}{\mathscr{B}}
\newcommand{\scrX}{\mathscr{X}}

\newcommand{\F}{\Phi}
\newcommand{\Y}{\Psi}

\newcommand{\rhoo}[2]{\rho_{#1}(#2)}
\newcommand{\bs}{\backslash}
\newcommand{\T}{\top}
\newcommand{\rarrow}{\rightarrow}

\newcommand{\fa}{\forall}

\title{\LARGE\textbf{Sparsity-Constrained Linear Quadratic Regulation Problem:\\ Greedy Approach with Performance Guarantee}}
\author{Shumpei Nishida and Kunihisa Okano%
 \thanks{The authors are with the Graduate School of Science and Engineering,
 Ritsumeikan University,
 Shiga 525-8577 Japan.
 E-mails: {\footnotesize \{\texttt{re0158ff@ed.}, \texttt{kokano@fc.}\}%
 \texttt{ritsumei.ac.jp}}.}%
 \thanks{This work was supported by JSPS KAKENHI Grant Number 20K14763.}%
}

\begin{document}

\maketitle
\thispagestyle{empty}
\pagestyle{empty}

\begin{abstract}
We study a linear quadratic regulation problem with a constraint where the control input can be nonzero only at a limited number of times.
Given that this constraint leads to a combinational optimization problem, we adopt a greedy method to find a suboptimal solution.
To quantify the performance of the greedy algorithm, we employ two metrics that reflect the submodularity level of the objective function: The submodularity ratio and curvature.
We first present an explicit form of the optimal control input that is amenable to evaluating these metrics.
Subsequently, we establish bounds on the submodularity ratio and curvature, which enable us to offer a practical performance guarantee for the greedy algorithm.
The effectiveness of our guarantee is further demonstrated through numerical simulations.
\end{abstract}

\section{Introduction} \label{sec:intro}

A signal is termed sparse if most of its values are exactly zero.
In control systems, a sparse control input means the input takes zero---the controller can remain turned off for the majority of the operational duration. 
The design of sparse control signals has attracted much research attention due to its energy-saving potential.
For example, in the control of railcars, leveraging inertia to move without any external driving force is an effective strategy to reduce power consumption.
Moreover, sparse control plays an important role in networked control systems by minimizing network usage.
Such a reduction is crucial for battery-powered devices and facilitates efficient network sharing among multiple nodes.
The seminal paper \cite{Nagahara2016a} has introduced a framework termed maximum hands-off control: In this approach, the control input remains at zero for as long as possible under a constraint on the terminal state.
A number of subsequent studies \cite{Kumar2019, Ikeda2021, Iwata2023, Nagahara2023} building on this foundation have followed.

In the above literature, the authors have focused on maximizing sparsity, with limited attention to control performance.
As a result, maximum hands-off control may lead to undesirable transient phenomena.
The present paper aims to establish a methodology taking into account both transient performance and control input sparsity. Several papers have studied optimal control with a limited number of control actions. In \cite{Imer2006} and \cite{Shi2013}, the authors have examined a strategy that applies control inputs at the beginning of the control horizon, which is optimal when control objectives disregard the energy of the control input. For the case of including the quadratic form of the input, a suboptimal method by convex relaxation has been proposed \cite{Jiao2023}.

Aside from these approaches, the greedy algorithm has been recognized as a practical approach to determine the actuation timings \cite{Chamon2019}. In addition to its simplicity in implementation, a notable feature of the greedy algorithm is that it offers theoretical performance guarantees. While such guarantees were traditionally associated with submodular objective functions \cite{Nemhauser1978}, recent studies have been extending these guarantees to non-submodular functions \cite{Bian2017}, which are prevalent in control problems \cite{Kohara2020, Kyriakis2020, Guo2021, Tzoumas2021, Li2023, Vafaee2023}. %

This paper addresses the design of the optimal control input that minimizes the standard quadratic cost subject to a sparsity constraint.
A greedy algorithm is employed to obtain an approximate solution.
In line with \cite{Bian2017}, we investigate the degree of submodularity of the objective function through the submodularity ratio \cite{Das2011} and generalized curvature \cite{Bian2017}. To evaluate these metrics, we structure the objective function by solving a least squares problem, while the dynamic programming approach has been used in \cite{Shi2013, Jiao2023, Chamon2019}. Numerical examples demonstrate that our result provides a tighter guarantee than those found in \cite{Chamon2019}.

The rest of the paper is organized as follows.
In Section \ref{sec:problem formulation}, we formulate the considered problem as a cardinality-constrained optimization problem.
Next, we provide preliminary results on the set function maximization in Section \ref{sec:greedy}.
We then present the optimal control input for the LQR problem in Section \ref{sec:opt control} and provide a performance guarantee for the greedy algorithm in Section \ref{sec:guarantee}.
We illustrate our performance guarantee in Section \ref{sec:simulation}.
Finally, we conclude our results in Section \ref{sec:conclusion}.

\textbf{Notation:}
We denote the set of real numbers and non-negative integers as $\bbR$ and $\bbZ$, respectively.
The $n \times n$ identity matrix is denoted by $I_n$, and $\diag\{d_1,\ldots,d_n\}$ represents a diagonal matrix where the diagonal elements are $d_i$s.
For a matrix $A$, $[A]_{i,j}$ means the $(i,j)$ element of $A$.
The spectral norm of $A$ is denoted by $\|A\|$.
For a symmetric matrix $S \in \bbR^{n \times n}$, its eigenvalues 
arranged in descending order are given by $\lambda_1(S) \geq \cdots \geq \lambda_n(S)$.
The symbol $\otimes$ denotes the Kronecker product.
For a finite set $\scrX$, $|\scrX|$ and $2^{\scrX}$ denote the cardinality and the power set of $\scrX$, respectively.

\section{Problem formulation} \label{sec:problem formulation}

\begin{figure}[!t]
    \centering

    \includegraphics[width=50mm]{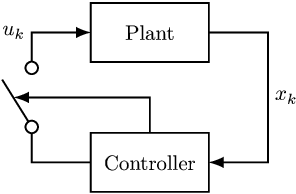}
    \caption{Feedback system}
    \label{fig:system}
\end{figure}

We investigate a discrete-time feedback system depicted in Fig.~\ref{fig:system}.
The plant is modeled as a linear time-invariant system given by
\begin{equation} \label{eq:system}
    x_{k+1} = Ax_k + Bu_k,
\end{equation}
where $x_k\in\mathbb{R}^n$ and $u_k\in\mathbb{R}^m$ are the state and the control input, respectively.
At each time step $k \in \bbZ$, the controller observes the state $x_k$ and determines the control input $u_k$.
Notably, the control input is a sparse signal, taking non-zero values at most $d$ times within an $N$-step control interval $\scrT \coloneqq \{0, 1, \ldots, N-1\}$. 

Let $\scrS \subseteq \scrT$ be the set of time instants at which the control input is allowed to be a nonzero value.
The objective of control is formally stated as the following minimization problem.
\begin{mini}|l|
    {u_0, \dots, u_{N-1}, \scrS}{x_N^{\T}Q_Nx_N + \sum_{k=0}^{N-1} x_k^{\T}Q_kx_k + u_k^{\T}R_ku_k,}{\label{eq:problem0}}{}
    \addConstraint{u_k = 0\quad (\forall k \notin \scrS)}
    \addConstraint{|\scrS| \leq d.}
\end{mini}
Here, $Q_k \succeq 0$  for all $k \in \scrT \cup \{N\}$, and $R_k \succ 0$ for all $k \in \scrT$. 
Problem \eqref{eq:problem0} entails a joint design of the control inputs $u_0,\dots,u_{N-1}$ and the actuation timing $\scrS$.
Once a timing set $\scrS$ is established, the optimal control inputs are determined by a solution of the standard LQR problem. We define the optimal LQR cost as
\begin{align}
    J(\scrS) = &\min_{u_0(\scrS),\dots,u_{N-1}(\scrS)} \left[x_N^{\T}Q_Nx_N\right.\\
    & + \left.\sum_{k=0}^{N-1} \left(x_k^{\T}Q_kx_k + u_k(\scrS)^{\T}R_ku_k(\scrS)\right)\right], \label{eq:obj function}    
\end{align}
where we use the notation $u_k(\scrS)$ to emphasize that the control inputs must satisfy the sparsity constraint: $u_k=0$ when $k\not\in\scrS$.
Consequently, the original problem \eqref{eq:problem0} can be transformed into the subsequent combinatorial optimization problem:
\begin{mini}|l|
    {\scrS \subseteq \scrT}{J(\scrS),}{\label{prob:actuator scheduling}}{}
    \addConstraint{|\scrS| \leq d.}
\end{mini}

Given that an efficient method for this problem has yet to be established, we adopt the greedy algorithm as an approximation technique.

\section{Greedy algorithm and (non-)submodular function maximization} \label{sec:greedy}
In this section, we provide preliminary results for the optimization of set functions.
Let us consider the following problem:
\begin{maxi}|l|
    {\scrS \subseteq \scrT}{f(\scrS),}{\label{prob:set function optimization problem}}{}
    \addConstraint{|\scrS| \leq d,}
\end{maxi}
where $f \colon 2^{\scrT} \rarrow \bbR$ is a set function.
The brute-force search over the feasible solutions becomes quickly intractable even for moderately sized problems.

The greedy algorithm, as shown in Algorithm~\ref{algo:greedy algorithm}, is one of the most common approximation methods for the aforementioned problem.
\begin{figure}[tb]
  \begin{algorithm}[H]
    \caption{The greedy algorithm for Problem~\eqref{prob:set function optimization problem}.}
    \label{algo:greedy algorithm}
    \begin{algorithmic}
        \renewcommand{\algorithmicrequire}{\textbf{Input:}}
        \renewcommand{\algorithmicensure}{\textbf{Output:}}
        \Require  Finite discrete set $\scrT$, set function $f$, integer $d$
        \State $\scrS_0 \leftarrow \emptyset$
        \For{$i=1,\cdots,d$}
            \State $\omega^* \leftarrow \underset{\omega \in \scrT\bs\scrS_{i-1}}{\text{arg\,max}}~f(\scrS_{i-1} \cup \{\omega\}) - f(\scrS_{i-1})$
            \State $\scrS_i \leftarrow \scrS_{i-1} \cup \{\omega^*\}$
        \EndFor
        \Ensure $\scrS^g \leftarrow \scrS_d$
    \end{algorithmic}
  \end{algorithm}
\end{figure}
Algorithm~\ref{algo:greedy algorithm} can yield a solution in polynomial time, which often performs well empirically.
Moreover, it is important to note that bounds exist on the deviation of greedy solutions from the optimal.

To describe a performance guarantee for the greedy algorithm, we now introduce fundamental notions associated with a set function.
Let us denote the marginal gain of a set $\Omega \subseteq \scrT$ with respect to a set $\scrS \subseteq \scrT$ by 
\begin{align}
    \rhoo{\Omega}{\scrS} \coloneqq f(\scrS \cup \Omega) - f(\scrS).
\end{align}

\begin{definition} \label{def:monotone nondecreasing}
    A set function $f\colon 2^{\scrT} \rarrow \bbR$ is \emph{monotone nondecreasing} if for all subsets $\scrA, \scrB$ that satisfy $\scrA \subseteq \scrB \subseteq \scrT$, it holds that
    \begin{equation}
        f(\scrA) \leq f(\scrB). \label{eq:monotone nondecreasing}
    \end{equation}
\end{definition}
\medskip

\begin{definition}[\!{\cite{Das2011}}] \label{def:submodularity ratio}
    The \emph{submodularity ratio} of a nonnegative set function $f$ is the largest scalar $\gamma$ such that
    \begin{equation}
        \sum_{\omega \in \Omega \bs \scrS} \rhoo{\{\omega\}}{\scrS} \geq \gamma \rhoo{\Omega}{\scrS},\quad \fa\Omega,\scrS \subseteq \scrT. \label{eq:submodularity ratio}
    \end{equation}
\end{definition}
\medskip

\begin{definition}[\!{\cite{Bian2017}}] \label{def:curvature}
    The \emph{curvature} of a nonnegative set function $f$ is the smallest scalar $\alpha$ such that
    \begin{align}
        &\rhoo{\{j\}}{\scrS\bs\{j\}\cup\Omega} \geq (1 - \alpha)\rhoo{\{j\}}{\scrS\bs\{j\}},\\
        &\fa \Omega,\scrS \subseteq \scrT,\quad \fa j \in \scrS\bs\Omega. \label{eq:curvature}
    \end{align}
\end{definition}
\medskip

For a nondecreasing set function, it holds that $\gamma \in [0,1]$ and $\alpha \in [0,1] $\cite{Bian2017}.

Let $\scrS^g$ be the solution to Problem~\eqref{prob:set function optimization problem} obtained by Algorithm~\ref{algo:greedy algorithm}, and $\scrS^*$ be the optimal solution.
With $\gamma$ and $\alpha$, the greedy algorithm provides an approximation guarantee for Problem~\eqref{prob:set function optimization problem}.

\begin{proposition}[\!{\cite{Bian2017}}] \label{prop:guarantee of non-submodular function}
    Let $f$ be a monotone nondecreasing set function with submodularity $\gamma \in [0,1]$ and curvature $\alpha \in [0,1]$.
    Then, Algorithm~\ref{algo:greedy algorithm} enjoys the following approximation guarantee for solving Problem \eqref{prob:set function optimization problem}:
    \begin{equation}
    \label{eq: performance guarantee for non-submodular function}
        f(\scrS^g) - f(\emptyset) \geq \frac{1}{\alpha}\left(1 - e^{-\alpha\gamma}\right)(f(\scrS^*) - f(\emptyset)).
    \end{equation}
\end{proposition}
\vskip\baselineskip

Proposition~\ref{prop:guarantee of non-submodular function} generalizes a well-known result \cite{Nemhauser1978} available for submodular functions to non-submodular functions. For the case of $\gamma=1$ and $\alpha=1$, i.e., when $f$ is submodular, the coefficient in \eqref{eq: performance guarantee for non-submodular function} is equal to $1 - e^{-1}$, which corresponds to the classical approximation factor given in \cite{Nemhauser1978}.

Note that computing $\gamma$ and $\alpha$ directly by Definitions \ref{def:submodularity ratio} and \ref{def:curvature} is intractable because the constraints in \eqref{eq:submodularity ratio} and \eqref{eq:curvature} are combinational.
Therefore, we consider deriving bounds on $\gamma$ and $\alpha$ and using them to establish a guarantee.
In the following section, we derive an explicit form of the optimal control input for Problem~\eqref{eq:problem0} and rewrite the objective function $J(\scrS)$ to evaluate its submodularity ratio and curvature.

\begin{remark}
    Recently, Harshaw \emph{et al.}\cite{Harshaw2019} have shown that no polynomial algorithm can achieve a better performance guarantee than \eqref{eq: performance guarantee for non-submodular function} for a nonnegative non-submodular function maximization with a cardinality constraint.
    Thus, we use Proposition \ref{prop:guarantee of non-submodular function} as an approximation guarantee for the greedy algorithm.
\end{remark}

\section{Explicit form of the optimal control input and the optimal cost} \label{sec:opt control}

We here present an explicit form of the optimal control in preparation for deriving our main result.
Let us define $\tilde{S}(\scrS) \in \bbR^{|\scrS| \times N}$ as the matrix created by removing the $i$th rows that hold $u_{i-1} = 0$ from $I_N$ for all $i = 1,\ldots,N$.
For a given $\scrS$, let $t_1 < t_2 < \cdots < t_{|\scrS|}$ be the elements of $\scrS$ in ascending order.
Then, $\tilde{S}$ is formally defined as follows:
\begin{equation} \label{eq:component of scheduling matrix}
    \left[\tilde{S}(\scrS)\right]_{i,j} = \begin{cases}
        1 & j-1=t_i,\\
        0 & \text{otherwise.}
    \end{cases}
\end{equation}
The number of rows of $\tilde{S}(\scrS)$ implies how many control inputs are allowed to be nonzero.
Using $\tilde{S}(\scrS)$, we define the matrix $S(\scrS) \in \bbR^{|\scrS|m \times Nm}$ as follows:
\begin{equation} \label{eq:scheduling matrix}
    S(\scrS) \coloneqq \tilde{S}(\scrS) \otimes I_m.
\end{equation}
In addition, let $I_N^{(i)}$ be the $N \times N$ matrix where $[I_N^{(i)}]_{i,i} = 1$ and the other elements are zero.
For simplicity of notation, we write $\tilde{S}(\scrS)$ and $S(\scrS)$ as $\tilde{S}$ and $S$ respectively in the subsequent text.

The following lemma holds.

\begin{lemma} \label{lemma:1}    
    Given a set $\scrS \subseteq \scrT$, it holds that
    \begin{equation}
        S^{\T}S = \sum_{i \in \scrS} \left(I_N^{(i+1)} \otimes I_m\right).
    \end{equation}
\end{lemma}
\medskip

\begin{proof}
    For a given set $\scrS$, $[\tilde{S}]_{i,t_i+1} = 1$ and the other elements are zero because of the definition of $\tilde{S}$.
    Then, we have that $[\tilde{S}^{\T}\tilde{S}]_{t_i+1,t_i+1} = 1$ for all $i=1,\dots,|\scrS|$ and the other elements are zero.
    Hence, it follows that
    \begin{align}
        S^{\T}S &= \left(\tilde{S} \otimes I_m\right)^{\T} \left(\tilde{S} \otimes I_m\right) = \left(\tilde{S}^{\T} \otimes I_m\right) \left(\tilde{S} \otimes I_m\right)\\
        &= \left(\tilde{S}^{\T}\tilde{S} \otimes I_m\right) = \left(\sum_{i=1}^{|\scrS|} I_N^{(t_i)} \otimes I_m\right)\\
        &= \sum_{i' \in \scrS} \left(I_N^{(i'+1)} \otimes I_m\right),
    \end{align}
    where $i' \coloneqq t_i - 1$. Here, the second, third, and last equalities hold from the properties of the Kronecker product.
\end{proof}
\medskip

Let $U \coloneqq [u_0(\scrS)^{\T},\ldots,u_{N-1}(\scrS)^{\T}]^{\T}$ be a control input  associated with a feasible set $\scrS$ satisfying the constraint in Problem~\eqref{eq:problem0}.
Then, the possibly nonzero inputs in $U$ are given as $SU = [u_{t_1}^{\T},\dots,u_{t_{|\scrS|}}^{\T}]^{\T}$.
Let also $U^* = [u_0^*(\scrS)^{\T},\dots,u_{N-1}^*(\scrS)^{\T}]^{\T}$ be the optimal control giving \eqref{eq:obj function}.

The following proposition shows the possibly nonzero values of the optimal input $SU^*$  and the optimal cost $J(\scrS)$ associated with $\scrS$.

\begin{proposition} \label{prop:opt control}
    Given a set $\scrS \subseteq \scrT$, it holds that
    \begin{equation}
        SU^* = -(S\bar{R}S^{\T} + S\bar{B}\F^{\top}\bar{Q}\F\bar{B}S^{\T})^{-1}S\bar{B}^{\T}\F^{\T}\bar{Q}\Y x_0,
    \end{equation}
    where
    \begin{equation}
    \label{eq:Phi,Psi}
        \begin{split}
            &\bar{Q} = \diag\{Q_1,\cdots,Q_N\},\quad \bar{R} = \diag\{R_0,\cdots,R_{N-1}\},\\
            &\F = \begin{bmatrix}
                I_n & 0 & \cdots & 0\\
                A & I_n & \cdots & 0\\
                \vdots & \vdots & \ddots & \vdots\\
                A^{N-1} & A^{N-2} & \cdots & I_n
            \end{bmatrix},\quad
            \Y = \begin{bmatrix}
                A\\
                A^2\\
                \vdots\\
                A^N
            \end{bmatrix},\\
            &\bar{B} = I_N \otimes B.
            \end{split}
    \end{equation}
    Furthermore, the optimal cost $J(\scrS)$ is given by
    \begin{equation}
        J(\scrS) = \tr\left[L(I_{Nn} +K(\scrS))^{-1}\right] + c, \label{eq:explicit J}
    \end{equation}
    where
    \begin{align}
        &L = \bar{Q}^{1/2}\Y x_0x_0^{\T}\Y^{\T}\bar{Q}^{1/2}, \label{eq:L} \\
        &K(\scrS) = \bar{Q}^{1/2}\F\bar{B}S^{\T}S\bar{R}^{-1}S^{\T}S\bar{B}^{\T}\F^{\T}\bar{Q}^{1/2}, \label{eq:K} \\
        &c = x_0^{\T}Q_0x_0. \label{eq:c}
    \end{align}
\end{proposition}
\medskip

\begin{remark}
    We emphasize that within the trace operator of the optimal cost expression \eqref{eq:explicit J}, a positive semidefinite matrix $L$ is present.
    This inclusion introduces a technical complexity when assessing the submodularity ratio and curvature of $J$.
    It is noted that in the existing work \cite{Kohara2020}, which also utilizes Proposition \ref{prop:guarantee of non-submodular function}, the objective function is described solely in terms of the inverse of a positive definite matrix.
\end{remark}
\medskip

Before proceeding to the proof of Proposition~\ref{prop:opt control}, we provide some lemmas in \cite{Bernstein2019}, which are used in the proof.

\begin{lemma} \label{lemma:2}
    Let $A \in \bbR^{n \times n}$ be symmetric matrix, and $P \in \bbR^{m \times n}$. If $\rank P = m$, then $PAP^{\T} \succ 0$.
\end{lemma}
\medskip

\begin{lemma} \label{lemma:square root matrix}
    Let $A$ be a positive semidefinite matrix, then there is a unique positive semidefinite square root matrix $A^{1/2}$ such that $A = A^{1/2}A^{1/2}$.
\end{lemma}
\medskip

\begin{lemma} \label{lemma:woodbury}
    Let $A \in \bbR^{n \times n}, B \in \bbR^{n \times m}, C \in \bbR^{m \times n}, D \in \bbR^{m \times m}$. If $A, D-CA^{-1}B$, and $D$ are nonsingular, then $A-BD^{-1}C$ is nonsingular and given by
    \begin{equation}
        (A - BD^{-1}C)^{-1} = A^{-1} - A^{-1}B(D - CA^{-1}B)^{-1}CA^{-1}.
    \end{equation}
\end{lemma}
\medskip

\par\noindent\hspace{2em}\emph{Proof of Proposition~\ref{prop:opt control}: }
    The solution of \eqref{eq:system} at time $k$ is given by
    \begin{equation} \label{eq:solution of system}
        x_k = A^kx_0 + \sum_{i=0}^{k-1} A^{k-i+1}Bu_i(\scrS),\quad \fa k \in\scrT \cup \{N\}.
    \end{equation}
    Here, let $X \coloneqq [x_1^{\T},\ldots,x_N^{\T}]^{\T}$.
    Combining \eqref{eq:solution of system} for $k=1,\ldots,N$, we have
    \begin{equation}
        X = \F\bar{B}U +\Y x_0, \label{eq:entire trajectory}
    \end{equation}
    where $\F$, $\bar{B}$, and $\Y$ are given by \eqref{eq:Phi,Psi}.
    From Lemma~\ref{lemma:1}, it follows that
    \begin{equation}
        U = S^{\T}SU.
    \end{equation}
    Thus, we can rewrite \eqref{eq:entire trajectory} as
    \begin{equation} \label{eq:X=}
        X = \F\bar{B}S^{\T}SU + \Y x_0.
    \end{equation}

    Let us define $V(u_0,\dots,u_{N-1}, \scrS)$ as
    \begin{align}
        &V(u_0,\dots,u_{N-1},\scrS)\\
        &\coloneqq x_N^{\T}Q_Nx_N + \sum_{k=0}^{N-1} \left(x_k^{\T}Q_kx_k + u_k(\scrS)^{\T}R_ku_k(\scrS)\right)\\
        &= X^{\T}\bar{Q}X + \left(S^{\T}SU\right)^{\T}\bar{R}\left(S^{\T}SU\right) + x_0^{\T}Q_0x_0. \label{eq:def of V}
    \end{align}
    By substituting \eqref{eq:X=} into \eqref{eq:def of V}, we have
    \begin{align}
        &V(u_0,\dots,u_{N-1},\scrS)\\
        &= \left(\F\bar{B}S^{\T}SU + \Y x_0\right)^{\T}\bar{Q}\left(\F\bar{B}S^{\T}SU + \Y x_0\right)\\
        &\quad + (SU)^{\T}S\bar{R}S^{\T}(SU) + x_0^{\T}Q_0x_0\\
        &= (SU)^{\T}\left(S\bar{R}S^{\T} + S\bar{B}^{\T}\F^{\T}\bar{Q}\F\bar{B}S^{\T}\right)(SU)\\
        &\quad + 2x_0^{\T}\Y^{\T}\bar{Q}\F\bar{B}S^{\T}(SU) + x_0^{\T}\Y^{\T}\bar{Q}\Y x_0 + x_0^{\T}Q_0x_0\\
        &= (SU + p)^{\T}M(SU + p) + c', \label{eq:V}
    \end{align}
    where
    \begin{align}
        &M = S\bar{R}S^{\T} + S\bar{B}^{\T}\F^{\T}\bar{Q}\F\bar{B}S^{\T},\\
        &p = M^{-1}S\bar{B}^{\T}\F^{\T}\bar{Q}\Y x_0, \label{eq:p} \\
        &c'= x_0^{\T}\Y^{\T}\bar{Q}\Y x_0 + x_0^{\T}Q_0x_0 - p^{\T}Mp.
    \end{align}
    Note that $M$ is nonsingular because $S$ is row full rank,
    $S\bar{R}S^{\T} \succ 0$ from Lemma~\ref{lemma:2},
    and $S\bar{B}^{\T}\F^{\T}\bar{Q}\F\bar{B}S^{\T} \succeq 0$.
    From \eqref{eq:V} and \eqref{eq:p}, the optimal control input is given by
    \begin{align}
        SU^* &= -p \label{eq:opt SU}\\
        &= -\left(S\bar{R}S^{\T} + S\bar{B}\F^{\top}\bar{Q}\F\bar{B}S^{\T}\right)^{-1}S\bar{B}^{\T}\F^{\T}\bar{Q}\Y x_0.
    \end{align}

    Since we have
    \begin{align}
        J(\scrS) &= \min_{u_0,\dots,u_{N-1}} V(u_0,\dots,u_{N-1},\scrS)\\
        &= V(u_0^*,\dots,u_{N-1}^*,\scrS),
    \end{align}
    it holds from \eqref{eq:V} and \eqref{eq:opt SU} that
    \begin{align}
        &J(\scrS) = c'\\
        &= x_0^{\T}\Y^{T}\bar{Q}^{1/2} \left[I_{Nn} - \left(\bar{Q}^{1/2}\F^{\T}\right) \left(\bar{B}S^{\T}S\bar{R}^{-1/2}\right)\right.\\
        &\quad \times \left.\left(\bar{B}S^{\T}S\bar{R}^{-1/2}\right)^{\T} \left(\bar{Q}^{1/2}\F^{\T}\right)^{\T}\right]^{-1} \bar{Q}^{1/2}\Y x_0 + c.
    \end{align}
    Here, 
    the second equality follows by introducing $\bar{Q}^{1/2}$ satisfying
    $\bar{Q} = \bar{Q}^{1/2}\bar{Q}^{1/2}$ according to Lemma~\ref{lemma:square root matrix}.
    By using Lemma~\ref{lemma:woodbury}, we can rewrite $J(\scrS)$ as
    \begin{align}
        J(\scrS) &= x_0^{\T}\Y^{T}\bar{Q} \left[I_{Nn} + \bar{Q}^{1/2}\F\bar{B}S^{\T}\left(S\bar{R}S^{\T}\right)^{-1}\right.\\
        &\quad \times \left. S\bar{B}^{\T}\F^{\T}\bar{Q}^{1/2}\right]^{-1}\bar{Q}^{1/2}\Y x_0 + c\label{eq:J pre}
    \end{align}
    In the right-hand side, we have $(S\bar{R}S^{\T})^{-1} = S\bar{R}^{-1}S^{\T}$ since $\bar{R}$ is block diagonal, which concludes \eqref{eq:explicit J}.
    \hspace*{\fill}~\QED\par
\medskip

Finally, we give an important property of $K(\scrS)$ which is used to derive the main result shown in Section~\ref{sec:guarantee}.

\begin{lemma} \label{lemma:5}
    For any given $\scrS \subseteq \scrT$ and any $\omega \in \scrT\bs\scrS$, it holds that
    \begin{equation}
        K(\scrS \cup \{\omega\}) = K(\scrS) + K(\{\omega\}).
    \end{equation}
\end{lemma}
\medskip

\begin{proof}
    From Lemma \ref{lemma:1}, for any $\scrS$, $K(\scrS)$ can be written as follows:
    \begin{align}
        K(\scrS) &= \bar{Q}^{1/2}\F\bar{B}S^{\T}S\bar{R}^{-1}S^{\T}S\bar{B}^{\T}\F^{\T}\bar{Q}^{1/2}\\
        &= \bar{Q}^{1/2}\F\bar{B} \sum_{i \in \scrS} \left(I_N^{(i+1)} \otimes I_m\right) \sum_{i=0}^{N-1} \left(I_N^{(i+1)} \otimes R_i^{-1}\right)\\
        &\quad \times \sum_{i \in \scrS} \left(I_N^{(i+1)} \otimes I_m\right) \bar{B}^{\T}\F^{\T}\bar{Q}^{1/2}\\
        &= \bar{Q}^{1/2}\F\bar{B} \sum_{i \in \scrS} \left(I_N^{(i+1)} \otimes I_m\right) \left(I_N^{(i+1)} \otimes R_i^{-1}\right)\\
        &\quad \times \left(I_N^{(i+1)} \otimes I_m\right) \bar{B}^{\T}\F^{\T}\bar{Q}^{1/2}\\
        &= \bar{Q}^{1/2}\F\bar{B} \sum_{i \in \scrS} \left(I_N^{(i+1)} \otimes R_i^{-1}\right) \bar{B}^{\T}\F^{\T}\bar{Q}^{1/2}.
    \end{align}
    In light of the far right-hand side of the above equations, we see that $K(\scrS)$ can be decomposed to the sum of positive semidefinite matrices corresponding to $t_1,\dots,t_{|\scrS|}\in\scrS$.
    This implies that
    \begin{equation}
        K(\scrS) = \sum_{\omega \in \scrS} K(\{\omega\})
    \end{equation}
    for any $\scrS \in \scrT$.
    Hence, for any fixed $\scrS \subseteq \scrT$ and any $\omega \in \scrT\bs\scrS$, we have
    \begin{align}
        K(\scrS \cup \{\omega\}) &= \sum_{\omega' \in \scrS \cup \{\omega\}} K(\{\omega'\})\\
        &= \sum_{\omega' \in \scrS} K(\{\omega'\}) + K(\{\omega\})\\
        &= K(\scrS) + K(\{\omega\}),
    \end{align}
    which concludes the proof.
\end{proof}

\section{Performance guarantee of the greedy algorithm} \label{sec:guarantee}
We are now ready to present our main result; a performance guarantee for the greedy algorithm to the optimal control with a sparsity constraint.
To apply Proposition \ref{prop:guarantee of non-submodular function} to Problem \eqref{prob:actuator scheduling}, we structure the problem as the following maximization:
\begin{maxi}|l|
    {\scrS \subseteq \scrT}{f(\scrS) \coloneqq -J(\scrS) + J(\emptyset),}{\label{prob:maximization problem}}{}
    \addConstraint{|\scrS| \leq d.}
\end{maxi}
Note that the sign of the objective function is flipped and $f(\emptyset) = 0$.
We seek to find a guarantee of the greedy solution to Problem~\eqref{prob:maximization problem} via Proposition \ref{prop:guarantee of non-submodular function}.
As already noted in Section~\ref{sec:greedy}, it is computationally difficult to find the exact values of $\gamma$ and $\alpha$.
Therefore, our goal is to bound them with computationally feasible values.

Let us define $\underline{\gamma}$ and $\overline{\alpha}$ as follows:
\begin{align}
    &\underline{\gamma} \!\coloneqq\! \frac{\min_{\omega \in \scrT} \tr[LK(\{\omega\})] \{\min_{\omega \in \scrT} \lambda_n[I_{Nn} + K(\{\omega\})]\}^2}{\max_{\omega \in \scrT} \tr[LK(\{\omega\})] \{\lambda_1[I_{Nn} + K(\scrT)]\}^2}, \label{eq:lb of gamma} \\
    &\overline{\alpha} \coloneqq 1 - \underline{\gamma}. \label{eq:ub of alpha}
\end{align}

\begin{remark}
    The values $\underline{\gamma}$ and $\overline{\alpha}$ are well defined if $\max_{\omega \in \scrT} \tr[LK(\{\omega\})] \neq 0$.
    Since all eigenvalues of $LK(\omega)$ are nonnegative for every $\omega\in\scrT$, this condition is typically met except in uninteresting cases such as $Ax_0=0$.
\end{remark}
\medskip

The above values represent bounds on $\gamma$ and $\alpha$ of $f(\scrS)$ in Problem \eqref{prob:maximization problem}.

\begin{theorem} \label{th:performance guarantee}
    Suppose that there exists $\omega \in \scrT$ such that $\tr[LK(\{\omega\})] \neq 0$.
    It holds that the set function $f(\scrS)$ defined in \eqref{prob:maximization problem} is monotone nondecreasing.
    In addition, its submodularity ratio $\gamma$ and curvature $\alpha$ are bounded by
    \begin{equation}
        1 \geq \gamma \geq \underline{\gamma} \geq 0,\quad 0 \leq \alpha \leq \overline{\alpha} \leq 1.
    \end{equation}
\end{theorem}
\medskip

To establish this theorem, we present preliminary lemmas from \cite{Bernstein2019}.

\begin{lemma} \label{lemma:inverse equality of positive define}
    For any positive definite matrices $A, B$, if $A \prec B$ then $A^{-1} \succ B^{-1}$.
\end{lemma}
\medskip

\begin{lemma} \label{lemma:trace inequality}
    Let $A, B$, and $C$ be symmetric matrices.
    If $A \succeq B$ and $C$ is positive semidefinite, then
    \begin{equation}
        \tr[CA] \geq \tr[CB].
    \end{equation}
\end{lemma}
\medskip

\begin{lemma} \label{lemma:eigen value of inverse}
    Let $A$ be a nonsingular matrix and $\lambda$ be an eigenvalue of $A$.
    Then, $1/\lambda$ is an eigenvalue of $A^{-1}$.
\end{lemma}
\medskip

\begin{lemma} \label{lemma:trace bounds}
    Let $A, B \in \bbR^{n \times n}$, assume that $A$ is positive semidefinite and $B$ is symmetric.
    Then it holds that
    \begin{equation}
        \sum_{i=1}^n \lambda_i[A]\lambda_{n-i+1}[B] \leq \tr[AB] \leq \sum_{i=1}^n \lambda_i[A]\lambda_i[B].
    \end{equation}
\end{lemma}
\medskip

\par\noindent\hspace{2em}\emph{Proof of Theorem~\ref{th:performance guarantee}: }
    The assumption that $\exists \omega : \tr\qty[LK(\omega)]\neq 0$ is required to guarantee $\underline{\gamma}$ and $\overline{\alpha}$ are well defined.
    Throughout this proof, we write $I_{Nn}$ as $I$ and
    \begin{equation}
        Z(\scrS) = I + K(\scrS).
    \end{equation}
    First, we show that $f(\scrS)$ is monotone nondecreasing.
    From Lemma \ref{lemma:5}, we have $ K(\scrS_1) \preceq K(\scrS_2)$ for all subsets $\scrS_1 \subseteq \scrS_2 \subseteq \scrT$.
    With this relation, it holds that $I + K(\scrS_1) \preceq I + K(\scrS_2)$.
    Combining this inequality and Lemma \ref{lemma:inverse equality of positive define}, we have that
    \begin{equation}
        (I + K(\scrS_1))^{-1} \succeq (I + K(\scrS_2))^{-1}.
    \end{equation}

    We note that $L$ is positive semidefinite because $L$ can be decomposed as \eqref{eq:L}.
    This implies that
    \begin{equation}
        L(I + K(\scrS_1))^{-1} \succeq L(I + K(\scrS_2))^{-1}.
    \end{equation}
   Then, we have
   \begin{equation}
       \tr\left[L(I + K(\scrS_1))^{-1}\right] \geq \tr\left[L(I + K(\scrS_2))^{-1}\right].
   \end{equation}
   from  Lemma \ref{lemma:trace inequality}.
   Hence, $J(\scrS)$ satisfies
    \begin{align}
        J(\scrS_1) &= \tr\left[L(I + K(\scrS_1))^{-1}\right] + c\\
        &\geq \tr\left[L(I + K(\scrS_2))^{-1}\right] + c\\
        &= J(\scrS_2).
    \end{align}
    for all subsets $\scrS_1 \subseteq \scrS_2 \subseteq \scrT$.
    Accordingly, we have that $f(\scrS_1) \leq f(\scrS_2)$, i.e., $f(\scrS)$ is monotone nondecreasing with respect to $\scrS$.

    Next, we derive the lower bound on the submodularity ratio $\gamma$.
    We begin by providing a lower bound on the left-hand side of \eqref{eq:submodularity ratio}:
    \begin{align}
        &\sum_{\omega \in \Omega\bs\scrS} \rhoo{\{\omega\}}{\scrS}\\
        &= \sum_{\omega \in \Omega\bs\scrS} -\tr\left[LZ(\scrS \cup \{\omega\})^{-1}\right] + \tr\left[LZ(\scrS)^{-1}\right]\\
        &\geq \sum_{\omega \in \Omega\bs\scrS} \sum_{i=1}^{Nn} \left\{ -\lambda_i[L]\lambda_i[Z(\scrS \cup \{\omega\})^{-1}] \right.\\
        &\quad\quad + \left. \lambda_i[L]\lambda_{Nn-i+1}[Z(\scrS)^{-1}] \right\}\\
        &= \sum_{\omega \in \Omega\bs\scrS} \sum_{i=1}^{Nn} \frac{1}{\lambda_i[Z(\scrS \cup \{\omega\})] \lambda_{Nn-i+1}[Z(\scrS)]}\\
        &\quad\quad \times \lambda_i[L] \left\{ \lambda_i[Z(\scrS \cup \{\omega\})] - \lambda_{Nn-i+1}[Z(\scrS)]\right\}\\
        &\geq \frac{\sum_{\omega \in \Omega\bs\scrS} \tr[LZ(\scrS \cup \{\omega\})] - \tr[LZ(\scrS)]}{\{\lambda_1[Z(\scrT)]\}^2}\\
        &= \frac{\sum_{\omega \in \Omega\bs\scrS} \tr[LK(\{\omega\})]}{\{\lambda_1[Z(\scrT)]\}^2}\\
        &\geq |\Omega\bs\scrS| \frac{\min_{\omega \in \scrT} \tr[LK(\{\omega\})]}{\{\lambda_1[Z(\scrT)]\}^2}. \label{eq:lb left-hand side gamma}
    \end{align}
    Here, the first and second inequalities are followed from Lemma \ref{lemma:trace bounds}.
    To obtain the second equality, we have used Lemma \ref{lemma:eigen value of inverse}. 
    The third equality holds by the fact that $K(\scrS \cup \{\omega\}) - K(\scrS) = K(\{\omega\})$, which is followed by Lemma \ref{lemma:5}.

    As to an upper bound on the right-hand side of \eqref{eq:curvature}, we have
    \begin{align}
        &\rhoo{\Omega}{\scrS}\\
        &= -\tr\left[LZ(\scrS \cup \Omega)^{-1}\right] + \tr\left[LZ(\scrS)^{-1}\right]\\
        &\leq\sum_{i=1}^{Nn} \left\{ -\lambda_i[L]\lambda_{Nn-i+1}\left[Z(\scrS \cup \Omega)^{-1}\right] \right.\\
        &\quad\quad \left. + \lambda_i[L]\lambda_i\left[Z(\scrS)^{-1}\right] \right\}\\
        &= \sum_{i=1}^{Nn} \frac{\lambda_i[L] \left\{ \lambda_{Nn-i+1}[Z(\scrS \cup \Omega)] - \lambda_i[Z(\scrS)]\right\}}{ \lambda_{Nn-i+1}[Z(\scrS \cup \Omega)] \lambda_i[Z(\scrS)]}\\
        &\leq \frac{\tr[LZ(\scrS \cup \Omega)] - \tr[LZ(\scrS)]}{\left\{ \min_{\omega \in \scrT} \lambda_n[Z(\{\omega\})] \right\}^2}\\
        &= \frac{\tr[LK(\Omega)]}{\left\{ \min_{\omega \in \scrT} \lambda_n[Z(\{\omega\})] \right\}^2}\\
        &\leq |\Omega\bs\scrS| \frac{\max_{\omega \in \scrT} \tr[LK(\{\omega\})]}{\left\{ \min_{\omega \in \scrT} \lambda_n[Z(\{\omega\})] \right\}^2}. \label{eq:ub right-hand side gamma}
    \end{align}
    We again used Lemma \ref{lemma:trace bounds} to obtain the first and the second inequalities.
    The second equality is also followed by Lemma~\ref{lemma:eigen value of inverse}.
    From \eqref{eq:lb left-hand side gamma} and \eqref{eq:ub right-hand side gamma}, we have
    \begin{align}
        \gamma
        &\geq \frac{\min_{\omega \in \scrT} \tr[LK(\{\omega\})] \left\{ \min_{\omega \in \scrT} \lambda_n[Z(\{\omega\})] \right\}^2}{\max_{\omega \in \scrT} \tr[LK(\{\omega\})] \{\lambda_1[Z(\scrT)]\}^2}\\
        &=\underline{\gamma}.      
    \end{align}

    We next derive an upper bound on the curvature via a similar discussion:
    A lower bound on the left-hand side of \eqref{eq:curvature} is given as
    \begin{align}
        &\rhoo{\{j\}}{\scrS\bs\{j\} \cup \Omega}\\
        &=  -\tr\left[LZ(\scrS \cup \Omega)^{-1}\right] + \tr\left[LZ(\scrS\bs\{j\} \cup \Omega)^{-1}\right]\\
        &\geq \sum_{i=1}^{Nn} \left\{ -\lambda_i[L]\lambda_i[Z(\scrS \cup \Omega)^{-1}] \right.\\
        &\quad\quad + \left. \lambda_i[L]\lambda_{Nn-i+1}[Z(\scrS\bs\{j\} \cup \Omega)^{-1}] \right\}\\
        &= \sum_{i=1}^{Nn} \frac{\lambda_i[L] \left\{ \lambda_i[Z(\scrS \cup \Omega)] - \lambda_{Nn-i+1}[Z(\scrS\bs\{j\} \cup \Omega)]\right\}}{ \lambda_i[Z(\scrS \cup \Omega)] \lambda_{Nn-i+1}[Z(\scrS\bs\{j\} \cup \Omega)]}\\
        &\geq \frac{\tr[LZ(\scrS \cup \Omega)] - \tr[LZ(\scrS\bs\{j\} \cup \Omega)]}{\{\lambda_1[Z(\scrT)]\}^2}\\
        &= \frac{\tr[LK(\{j\})]}{\{\lambda_1[Z(\scrT)]\}^2}\\
        &\geq \frac{\min_{\omega \in \scrT} \tr[LK(\{\omega\})]}{\{\lambda_1[Z(\scrT)]\}^2}. \label{eq:lb left-hand side alpha}
    \end{align}
    Furthermore, we can obtain an upper bound on the right-hand side of \eqref{eq:curvature} as
    \begin{align}
        &\rhoo{\{j\}}{\scrS}\\
        &= -\tr\left[LZ(\scrS)^{-1}\right] + \tr\left[LZ(\scrS\bs\{j\})^{-1}\right]\\
        &\leq\sum_{i=1}^{Nn} \left\{ -\lambda_i[L]\lambda_{Nn-i+1}\left[Z(\scrS)^{-1}\right] \right.\\
        &\quad\quad \left. + \lambda_i[L]\lambda_i\left[Z(\scrS\bs\{j\})^{-1}\right] \right\}\\
        &= \sum_{i=1}^{Nn} \frac{\lambda_i[L] \left\{ \lambda_{Nn-i+1}[Z(\scrS)] - \lambda_i[Z(\scrS\bs\{j\})]\right\}}{ \lambda_{Nn-i+1}[Z(\scrS)] \lambda_i[Z(\scrS\bs\{j\})]}\\
        &\leq \frac{\tr[LZ(\scrS)] - \tr[LZ(\scrS\bs\{j\})]}{\left\{ \min_{\omega \in \scrT} \lambda_n[Z(\{j\})] \right\}^2}\\
        &= \frac{\tr[LK(\{j\})]}{\left\{ \min_{j \in \scrT} \lambda_n[Z(\{j\})] \right\}^2}\\
        &\leq \frac{\max_{\omega \in \scrT} \tr[LK(\{\omega\})]}{\left\{ \min_{\omega \in \scrT} \lambda_n[Z(\{\omega\})] \right\}^2}. \label{eq:ub right-hand side alpha}
    \end{align}
    From \eqref{eq:lb left-hand side alpha} and \eqref{eq:ub right-hand side alpha}, we conclude
    \begin{align}
        1 - \alpha
        &\geq \frac{\min_{\omega \in \scrT} \tr[LK(\{\omega\})] \left\{ \min_{\omega \in \scrT} \lambda_n[Z(\{\omega\})] \right\}^2}{\max_{\omega \in \scrT} \tr[LK(\{\omega\})] \{\lambda_1[Z(\scrT)]\}^2}\\
        &=1-\overline{\alpha}.
    \end{align}
    This implies the upper bound on the curvature as in \eqref{eq:ub of alpha}.

    Finally, $\underline{\gamma} \geq 0$ and $\overline{\alpha} \leq 1$ hold since $\min_{\omega \in \scrT} \tr[LK(\{\omega\})] \geq 0$ for all $\omega \in \scrT$.
    In addition, $1 \geq \gamma \geq 0$ and $0 \leq \alpha \leq 1$ hold because $f(\scrS)$ is monotone nondecreasing as described above.
    \hspace*{\fill}~\QED\par
\medskip

Let $\scrS^g$ be the solution to Problem~\eqref{prob:maximization problem} obtained by Algorithm \ref{algo:greedy algorithm}, and let $\scrS^*$ be the optimal solution.
From Theorem \ref{th:performance guarantee} and Proposition \ref{prop:guarantee of non-submodular function}, a performance guarantee of the greedy solution to the problem can be obtained as follows.

\begin{corollary}\label{cor:main result}%
    Consider the function $f(\scrS)$ in Problem \eqref{prob:maximization problem}
    and suppose that the assumption in Theorem \ref{th:performance guarantee} is satisfied.
    The following inequality holds:
    \begin{equation}
    \label{eq:lb factor}
        f(\scrS^g) \geq \frac{1}{\overline{\alpha}} \left(1 - e^{-\overline{\alpha}\underline{\gamma}}\right)f(\scrS^*).
    \end{equation}
\end{corollary}

\vskip\baselineskip

The above corollary provides a feasible approximation factor because $\underline{\gamma}$ and $\overline{\alpha}$ can be computed in polynomial time.
We use this corollary to provide the approximation guarantee with a numerical example in the next section. 

\section{Numerical examples and comparison with the existing results} \label{sec:simulation}
In this section, we validate the effectiveness of the greedy solution to Problem \eqref{prop:opt control} through a numerical example.
Furthermore, the conservativeness of the guarantee given in Corollary \ref{cor:main result} is discussed.

\subsection{Control performance of the greedy solution}

Consider the control of the plant described by \eqref{eq:system} with 
\begin{equation}
    A = \begin{bmatrix}
        1.1 & 1 & 0 & 0 & 0\\
        0 & 1.1 & 1 & 0 & 0\\
        0 & 0 & 1.1 & 1 & 0\\
        0 & 0 & 0 & 1.1 & 1\\
        0 & 0 & 0 & 0 & 1.1
    \end{bmatrix},\:
    B = 0.1I_5,\:
    x_0 = \begin{bmatrix}
        1\\
        1\\
        1\\
        1\\
        1
    \end{bmatrix}.
\end{equation}
The control horizon is set as $N=50$, and the parameters of the objective function in \eqref{eq:problem0} are given as $Q_k=0.1I_n$ and $R_k=I_n$ for all $k$.

\begin{figure}[t]
    \centering
    \includegraphics[width=80mm]{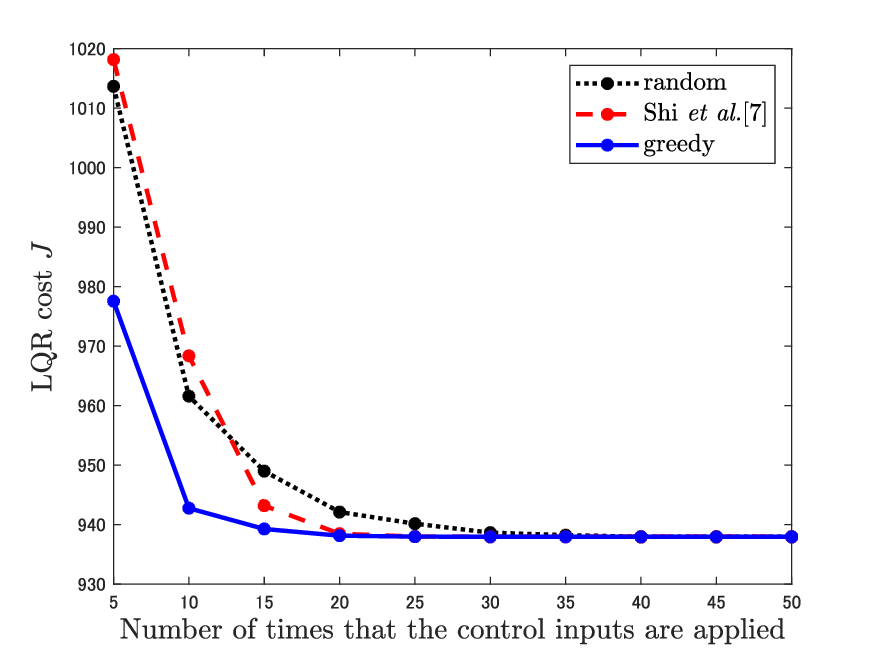}
    \caption{LQR cost defined by \eqref{eq:obj function} versus the number of times $d$ that the control inputs are applied.}
    \label{fig:cost}
\end{figure}

We investigate three methods to determine timing sets $\scrS$ and evaluate the corresponding costs $J(\scrS)$:
The first method involves randomly selecting time steps $\scrS$ subject to the sparsity constraint, and then computing the optimal control according to Proposition \ref{prop:opt control}. Among 1000 trials, the input that achieves the lowest cost is chosen.
The second method is adopted from \cite{Shi2013}, where the control inputs are applied during the first $d$ time steps.
The last one is Algorithm \ref{algo:greedy algorithm}---the greedy algorithm.
Fig.~\ref{fig:cost} compares the three cases by showing the values of $J(\scrS)$ versus the maximum number of control actions $d$.
We see that the greedy algorithm achieves lower costs than those by the random policy and \cite{Shi2013}, particularly when $d$ is smaller than 20.

It should be noted that the feedforward control is employed in this simulation: The control inputs for the entire horizon are computed at the initial time depending on $x_0$.
In practical systems, the input can be updated at each time instance based on the latest observation and the number of times that the input has been applied.

\subsection{Performance guarantee of the greedy algorithm} \label{subsec:B}

We now illustrate how the established performance guarantee vary depending on the target system.
In the following, we call $f(\scrS^g)/f(\scrS^*)$ the \emph{approximation ratio} of Algorithm \ref{algo:greedy algorithm} for Problem \eqref{prob:maximization problem}.
Since both $\underline{\gamma}$ and $\overline{\alpha}$ contain the powers of $A$, it is expected that singular values of $A$ characterize the approximation ratio.
Therefore, we examine the ratio in relation to the maximum singular value, which is the spectral norm of $A$.

\begin{figure}[t]
    \centering
    \includegraphics[width=80mm]{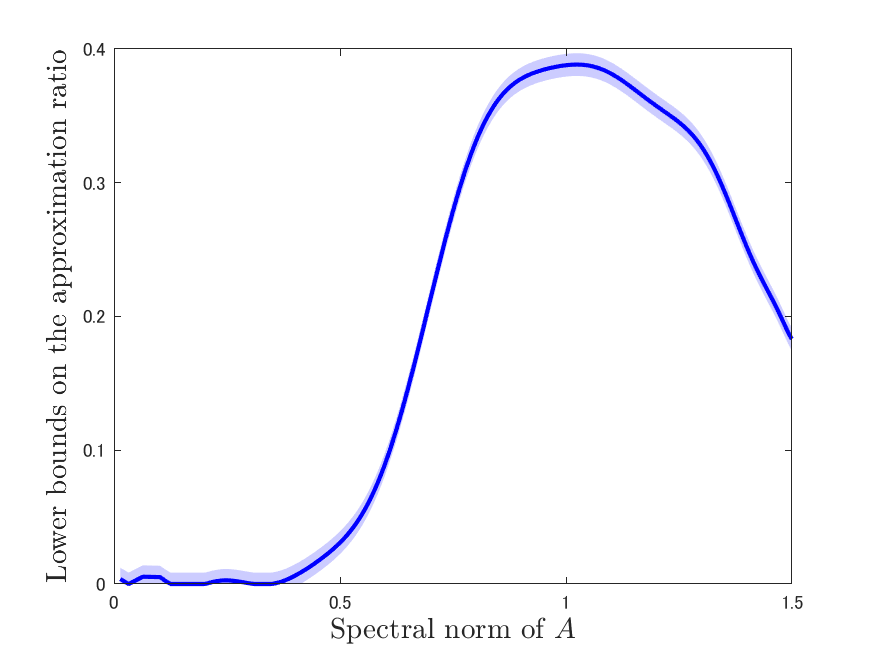}
    \caption{Lower bounds on the approximation ratio versus the spectral norm of $A$. The blue solid line represents the mean for 1000 simulations, and the shadowed region around the line visualizes standard deviations from the mean.}
    \label{fig:approximation ratio vs SN}
\end{figure}

Consider the system \eqref{eq:system} with $n=2$ and the control horizon of $N=5$.
Suppose that $n=m$ and $B = 0.1I_n$.
Furthermore, we assume that $Q_k = 0.1I_n$ for all $k \in \scrT \cup \{N\}$, $R_0=10I_n$, and $R_k = 10/k^2I_n$ for all $k \in \scrT\bs\{0\}$.
The matrix $A$ is randomly chosen as $A = \diag\{a_1, a_2\}$ where $|a_1|, |a_2| \leq 1.5$.
The initial state $x_0$ is also randomly chosen so that all elements are in $[-10,10]$.
Fig.~\ref{fig:approximation ratio vs SN} shows the lower bound on the approximation ratio given by Corollary \ref{cor:main result} in relation to $\|A\|$.
The solid line depicts the mean over 1000 realizations, while the shaded area represents the standard deviation.
It is observed that the approximation ratio archives about $0.4$ when $\|A\|$ is around 1.
On the other hand, the bound is markedly small when $\|A\|$ is closed to 0.1; the system is highly stable in those cases.
For the case where $\|A\| > 1$, the spectral radius of $A$ can be greater than 1, that is, a system is unstable.
The instability results in the magnification of $L$ and $K(\scrS)$, and thus $\max_{\omega \in \scrT} \tr[LK(\{\omega\})]$ becomes much greater than $\min_{\omega \in \scrT} \tr[LK(\{\omega\})]$.
Consequently, the approximation ratio tends to decrease as $\|A\|$ increases.

\subsection{Conservativeness of the established guarantee}

Finally, we analyze the conservativeness of Theorem~\ref{th:performance guarantee}.
The authors in \cite{Chamon2019} have explored an actuator scheduling problem, aiming to select a subset of the actuators to apply the control inputs at each time in order to minimize the control cost.
A linear deterministic system with a Gaussian initial state is considered.
The problem is formulated as a matroid-constrained optimization problem, and a performance guarantee of the greedy approach is derived using the concept of $\alpha$-supermodularity \cite{Lehmann2006}.
It is worth noting that \cite{Chamon2019} assumes that $A$ is full rank, which is required for applying the results of \cite{Chamon2017}.

Theorem \ref{th:performance guarantee} can be modified to the case where $x_0$ is a random vector.
In such a case, the optimal cost in Proposition \ref{prop:opt control} becomes the expectation of $J(\scrS)$, which is characterized by the covariance of $x_0$.
With these modifications, it is possible to compare the approximation ratio in Corollary \ref{cor:main result} with the result presented in \cite{Chamon2019}.
Suppose the same scenario as in Section~\ref{subsec:B} except here $x_0 \sim \mathcal{N}(0, I_n)$.
According to Corollary \ref{cor:main result}, we have that $f(\scrS^g)/f(\scrS^*) \leq 0.264$ on average over 1000 trials, whereas the result in \cite{Chamon2019} yields $f(\scrS^g)/f(\scrS^*) \leq 0.089$.
This comparison suggests that our result is less conservative.

\section{Conclusion} \label{sec:conclusion}

In this paper, we have addressed the sparsity-constrained LQR problem.
To evaluate the approximation guarantee of the greedy solution, we have initially derived the explicit optimal control input by solving the least squares problem.
With the form of the optimal control, bounds on the submodularity ratio and curvature of the quadratic cost function have been derived.
Those bounds have been utilized to establish a theoretical performance guarantee of the greedy solution.
Through numerical simulations, we have illustrated the effectiveness of the greedy solution and the performance guarantee.

\bibliography{my_biblib}
\end{document}